\documentclass[aps,prl,twocolumn,superscriptaddress,amsmath,amsfonts,showpacs]{revtex4}

\usepackage{graphicx}
\usepackage{bm} 
\begin{document}


\title{
Non-Abelian Topological Order in $S$-Wave Superfluids of
Ultracold Fermionic Atoms 
}


\author{Masatoshi Sato}
\affiliation{The Institute for Solid State Physics, The University of
Tokyo, Kashiwanoha 5-1-5, Kashiwa-shi, Chiba 277-8581, Japan}
\author{Yoshiro Takahashi}
\affiliation{Department of Physics, Kyoto University, Kyoto 606-8502, Japan}
\author{Satoshi Fujimoto}
\affiliation{Department of Physics, Kyoto University, Kyoto 606-8502, Japan}




\date{\today}

\begin{abstract}
It is proposed that in $s$-wave superfluids of cold fermionic atoms 
with laser-field-generated effective spin-orbit interactions, 
a topological phase with gapless edge states and 
Majorana fermion quasiparticles 
obeying non-Abelian statistics is realized in the case with a large
Zeeman magnetic field.
Our scenario provides a promising approach to the realization of
quantum computation based on the manipulation of non-Abelian anyons
via an $s$-wave Feshbach resonance.
\end{abstract}



\maketitle



{\it Introduction --}
Recently, there has been considerable interest in 
topological phases of quantum many-body systems, 
which are 
characterized by the following features 
\cite{wen,moore,nayak,zhang2,kane,zhang,kitaev,lee}: 
(i) there are topologically-protected 
gapless edge states
on surface boundaries of the systems, which are stable against
local perturbations, (ii) for two-dimensional (2D) systems, 
there are quasiparticles 
with fractional quantum numbers (e.g.,
fractional charges) 
termed ``anyons''. To this time,
the possibility of realizing topological phases has been studied
for various states realized in condensed matter systems,
such as quantum (spin) Hall states \cite{moore,nayak,zhang2,kane,zhang}, 
vortex states of $p+ip$ superconductors \cite{read,ivanov,stern2,machida},
and spin liquid states \cite{kitaev,zol}, and for cosmological systems
such as axion strings \cite{Sato03}.
The feature (ii) is particularly of interest 
in connection with the realization of fault-tolerant quantum 
computation based on the manipulation of non-Abelian
anyons\cite{kitaev,kitaev2,freedman,stone1,tewari2,zol}.
Since topological phases provide not only 
a novel paradigm of quantum ground states
but also a potential breakthrough for technological advance, 
it is desirable to pursue various possible schemes 
for their realization.

In this letter, we propose a scenario in which a topological phase,
possessing gapless edge states and 
non-Abelian anyons, is realized in a BCS $s$-wave superfluid (SF)
of ultracold fermionic atoms in an optical lattice
with a laser-field-generated effective spin-orbit (SO) interaction.
It is possible to generate an artificial SO
interaction that acts on atoms by using spatially varying laser 
fields \cite{ost,ruse,zhu,stan,LCPPPS09}.
The effective SO interaction is 
a key factor in our scenario of the topological phase.
Recently, topological phases in superconductors and SFs
have been investigated by several 
authors \cite{read,ivanov,stern2,machida,tewari2,Sato06,TYBN08,SF08,ZTLDS}.
The previous studies, however, 
focus on the $p$-wave pairing state 
\footnote{Also, in ref. \cite{ZTLDS}, the realization of 
a topological phase via an $s$-wave Feshbach resonance is proposed.
However, in ref.\cite{ZTLDS}, a $p$-wave SF stabilized through
an effective $p$-wave attractive interaction 
is considered, and the non-Abelian anyons are vortices of
the SF order parameter. 
Thus, the scenario considered in ref. \cite{ZTLDS} is distinctly  
different from that in the current paper.
This difference is also clearly seen from the fact that
the topological order in the current paper is due to
fermions with $k_F\sim 0$ as clarified below Eq.(\ref{eq:dual2}), 
while, by contrast,
in ref. \cite{ZTLDS}, topological order is due to fermions with
$k_F\sim \lambda/v_F$.
}.
BCS gaps of $p$-wave superconductors in solid state systems
are typically very small, and it is difficult to utilize them
for topological quantum computation, because
topological phases are destroyed by thermal excitations beyond bulk energy gaps.
For cold atoms, in principle, a $p$-wave SF with a large BCS gap can
be produced via a $p$-wave Feshbach resonance \cite{tewari2}. However,
unfortunately, this has not yet been  
realized because of huge loss \cite{inada}.
Contrastingly, 
$s$-wave SFs of cold atoms with large BCS gaps
have been realized via
an $s$-wave Feshbach resonance \cite{sbcs,chin}.
Thus, our scenario based on $s$-wave SFs of cold atoms
is deemed more advantageous 
for the realization of the topological order
than that using $p$-wave SFs via a $p$-wave Feshbach resonance.
Moreover, 
there is an important difference between 
the topological phase considered here
and that of $p$-wave SFs.
For chiral $p$-wave SFs, 
the non-Abelian anyons are vortices of
the SF order parameter, which contain Majorana fermion modes.
In striking contrast,
in our system, {\it the non-Abelian anyons are vortices of
the SO interaction, i.e. the phase twist caused by the orbital motion
accompanying spin flip.} 
We propose an experimental scheme for generating and controlling 
vortices in
the SO interaction, i.e. non-Abelian anyons,
which are stabilized by use of a carefully designed laser setup rather than
spontaneously formed macroscopic condensates. 
This scheme can be carried out by utilizing
current sophisticated laser techniques.
Thus, our proposal provides a promising approach to the realization of 
topological quantum
computation based on the manipulation of non-Abelian anyons.

{\it Model and Analysis of the topological phase --}
Let us consider an $s$-wave SF of neutral fermionic atoms in the 2D
optical square lattice, which is described by the Hamiltonian ${\cal H}={\cal H}_{\rm
kin}+{\cal H}_{\rm SO}+{\cal H}_{\rm s}$:
\begin{eqnarray}
{\cal H}_{\rm kin}
&=&
-t\sum_{{\bm i}\sigma} \sum_{\hat{\mu}=\hat{x},\hat{y}}(c_{{\bm i}+\hat{\mu}\sigma}^{\dagger}
 c_{{\bm i}\sigma}+c_{{\bm i}-\hat{\mu}\sigma}^{\dagger}c_{{\bm i}\sigma})
 -\mu\sum_{{\bm i}\sigma}c_{{\bm i}\sigma}^{\dagger}c_{{\bm i}\sigma}
\nonumber\\ 
&& -h\sum_{\bm i}
( c_{{\bm i}\uparrow}^{\dagger}c_{{\bm i}\uparrow}
-c_{{\bm i}\downarrow}^{\dagger}c_{{\bm i}\downarrow}),
\nonumber\\
{\cal H}_{\rm SO}&=&-\lambda \sum_{{\bm i}}[
(c^{\dagger}_{{\bm i}-\hat{x}\downarrow}
c_{{\bm i}\uparrow}-c^{\dagger}_{{\bm i}+\hat{x}\downarrow}
c_{{\bm i}\uparrow})
\nonumber\\
&&+i(c_{{\bm i}-\hat{y}\downarrow}^{\dagger}
c_{{\bm i}\uparrow}-c^{\dagger}_{{\bm i}+\hat{y}\downarrow}
c_{{\bm i}\uparrow})+{\rm h.c.}],
\nonumber\\
{\cal H}_{\rm s}&=&-\sum_{\bm i}\psi_{\rm s}
 (c_{{\bm i}\uparrow}^{\dagger}c_{{\bm i}\downarrow}^{\dagger}+{\rm h.c.}),
\label{eq:latticehamiltonian}
\end{eqnarray}
where $c^{\dagger}_{{\bm i}\sigma}$ ($c_{{\bm i}\sigma}$) denotes a
creation (annihilation) operator of the fermionic atom with pseudo-spin
$\sigma=(\uparrow,\downarrow)$ at site ${\bm i}=(i_x,i_y)$, and
$\psi_{\rm s}$ the gap function. $\hat{x}$ ($\hat{y}$) is a basic
lattice vector along the $x$($y$)-axis. 
${\cal H}_{\rm SO}$ is an effective Rashba type SO interaction \cite{rash}. 
We will discuss later the method of 
generating the Rashba SO interaction for neutral atoms via laser fields.
We also introduce the chemical potential
$\mu$ and the Zeeman term induced by a magnetic field $h$.
In the momentum space, the Hamiltonian is recast into
\begin{eqnarray}
{\cal H}=\frac{1}{2}\sum_{\bm k}
\left({\bm c}_{\bm k}^{\dagger}
, {\bm c}_{-{\bm k}}
\right)
{\cal H}({\bm k})
\left(
\begin{array}{c}
{\bm c}_{\bm k}\\
{\bm c}_{-{\bm k}}^{\dagger} \\
\end{array}
\right), 
\end{eqnarray}
with ${\bm c}^{\dagger}_{\bm k}=1/\sqrt{V}\sum_{\bm i}
e^{i{\bm k}{\bm i}}(c_{{\bm i}\uparrow}^{\dagger}, c_{{\bm i}\downarrow}^{\dagger})$,
and 
\begin{eqnarray}
{\cal H}({\bm k})=
\left(
\begin{array}{cc}
\epsilon_{\bm k}-h\sigma_z+{\bm g}_{\bm k}\cdot{\bm \sigma}
& i\psi_{\rm s} \sigma_y\\
-i\psi_{\rm s}\sigma_y
& -\epsilon_{\bm k}+h\sigma_z+{\bm g}_{\bm k}\cdot {\bm \sigma}^{*}
\end{array}
\right), 
\label{eq:khamiltonian}
\end{eqnarray}
where $\epsilon_{\bm k}=-2t(\cos k_x +\cos k_y)-\mu$,
${\bm g}_{\bm k}=2\lambda(\sin k_y, -\sin k_x)$,  
and $\bm \sigma=(\sigma_x,\sigma_y)$ 
the Pauli matrices.

As mentioned in the introduction, 
the non-Abelian topological order is characterized by the existence of
gapless chiral edge states propagating only in one direction 
and the existence of the non-Abelian anyons~\cite{read}.
The former is also associated with the nonzero Chern number~\cite{TKNN}.
In the following, we demonstrate that these features are 
indeed realized in the system (\ref{eq:latticehamiltonian}) when
a certain relation among
$\mu$, $h$, and $\psi_s$ holds. (Eq.(\ref{eq:region}) below)

A key observation of our analysis is that the Hamiltonian ${\cal
H}({\bm k})$ is unitary equivalent to the following ``dual''
Hamiltonian ${\cal H}^{\rm D}({\bm k})$,
\begin{eqnarray}
{\cal H}^{\rm D}({\bm k})=
\left(
\begin{array}{cc}
\psi_{\rm s}-h\sigma_z
&-i\epsilon_{\bm k}\sigma_y-i{\bm g}_{\bm k}\cdot{\bm \sigma}\sigma_y \\
i\epsilon_{\bm k}\sigma_y+i{\bm g}_{\bm k}\sigma_y{\bm \sigma}
& -\psi_{\rm s}+ h\sigma_z
\end{array}
\right), 
\label{eq:dualHamiltonian}
\end{eqnarray}
with the unitary transformation
\begin{eqnarray}
{\cal H}^{\rm D}({\bm k})=D{\cal H}({\bm k})D^{\dagger},
\quad
D=\frac{1}{\sqrt{2}}
\left(
\begin{array}{cc}
1 & i\sigma_y \\
i\sigma_y & 1
\end{array}
\right). 
\label{eq:duality}
\end{eqnarray}
From Eq.(\ref{eq:dualHamiltonian}), it is found that
the Rashba SO interaction ${\bm g}_{\bm k}\cdot{\bm \sigma}$ 
in the original Hamiltonian ${\cal H}({\bm k})$ 
is formally tranformed into 
a ``$p$-wave SF gap''
with the ${\bm d}$ vector, ${\bm
d}^{\rm D}_{\bm k}\equiv -{\bm g}_{\bm k}$, in
the dual Hamiltonian ${\cal H}^{\rm D}({\bm k})$.
However, this does not necessarily
mean that the topological properties of ${\cal H}({\bm k})$ are the same
as those of a $p$-wave SF, 
since ${\cal H}^{\rm D}({\bm
k})$ has a non-standard constant kinetic term 
$\epsilon^{\rm D}_{\bm k}\equiv\psi_{\rm s}$.
A similar $p$-wave SF state with a constant kinetic energy term was
considered before in the context of the quantum-Hall-effect (QHE) 
state~\cite{read}.
An important feature of (\ref{eq:dualHamiltonian})
is that the topological order emerges when $\mu$, $h$, and $\psi_s$ satisfy
\begin{eqnarray}
\psi_{\rm s}^2+\epsilon(0,0)^2<h^2<\psi_{\rm s}^2+\epsilon(\pi,0)^2, 
\label{eq:region}
\end{eqnarray}
with $\epsilon(k_x,k_y)\equiv \epsilon_{\bm k}$.
Here note that although the condition (\ref{eq:region}) implies
the Zeeman energy larger than the BCS gap $\psi_s$, the superfluidity is stable 
when $\lambda \gg h$ \cite{Frigeri,fuji3}. 
This stability is specific to neutral atomic systems. For electron systems,
such large magnetic fields usually destroy superconductivity via 
an orbital depairing effect.

\begin{figure}
\includegraphics[width=8cm]{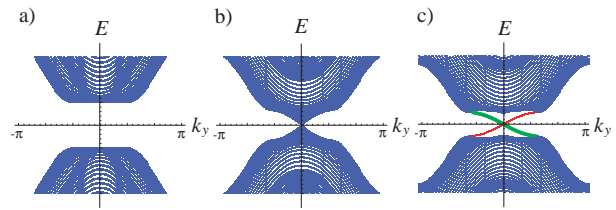}
\caption{\label{fig0} The band energy of the lattice Hamiltonian
 (\ref{eq:latticehamiltonian}) with edges at $i_x=0$ and $i_x=50(=L)$.
Here $k_y\in [-\pi,\pi]$ denotes the momentum in the $y$-direction. 
We set $t=1$, $\mu=-4$, $\lambda=0.5$ and $\psi=0.5$.
$h$ is a) $h=0$, b) $h=0.5$, c) $h=0.8$. 
The red thin line indicates a gapless chiral edge mode localized on the one side and
green thick line a gapless chiral edge mode on the other side.
They appear for $\sqrt{\psi^2+\epsilon(0,0)^2}
<h<\sqrt{\psi^2+\epsilon(0,\pi)^2}$.}
\end{figure}

Let us first examine edge states in our model.
Figure \ref{fig0} illustrates the energy bands obtained by diagonalizing the
lattice Hamiltonian (\ref{eq:latticehamiltonian}) with the open boundaries
at $i_x=0, L$ for various
$h$.
Here we have taken the periodic boundary condition in the $y$ direction,
and $k_y\in [-\pi,\pi]$ is the lattice momentum in the $y$ direction.
By increasing $h$ from zero adiabatically, it is found that the bulk
energy gap closes at 
$h=\sqrt{\psi_{\rm s}^2+\epsilon(0,0)^2}$ (Fig.\ref{fig0}(b)), 
then, for $h$ satisfying (\ref{eq:region}), 
a gapless edge mode with a linear dispersion 
$E\sim c k_y$ ($E\sim -c k_y$) localized on the one edge
(the other edge) appears between the bulk energy gap (Fig.\ref{fig0}(c)).
This chiral edge state is stable against any weak local perturbations provided that
there exists the nonzero 
Chern number; i.e. the topological number equivalent to 
the total number of gapless chiral edge modes, 
which was first introduced in the case of the QHE states~\cite{TKNN}.
We calculated the Chern number ${\cal Q}$ 
for $\mathcal{H}({\bm k})$ or equivalently $\mathcal{H}^{\rm D}({\bm k})$.
(Since the Chern number is calculated from the Berry curvature 
in the momentum space, it is not affect by the unitary transformation
$D$ which is independent of ${\bm k}$, ensuring 
the topological equivalence between (\ref{eq:khamiltonian})
and (\ref{eq:dualHamiltonian}).)
We found that ${\cal Q}=1$ 
when the condition (\ref{eq:region}) is satisfied~\cite{STF09-2}.
This is consistent with the numerical results for edge states shown above.

We, now, demonstrate that there exist the non-Abelian anyons in our system.
For this purpose,
we solve the Bogoliubov-de Gennes (BdG) equation for a single vortex:
If there exists a single Majorana fermion zero mode for each vortex, 
vortices obey the
non-Abelian statistics \cite{read,ivanov}.
We use the dual Hamiltonian ${\cal H}^{\rm D}$ to solve the BdG
equation, then construct a solution in the original Hamiltonian ${\cal
H}$ by using the duality transformation (\ref{eq:duality}).
For simplicity, we assume $\epsilon(0,0)=0$ for the time being.
Then, low-energy properties are governed by fermions on the Fermi
surface, which is split into $|{\bm k}|\sim 0$ and $|{\bm k}|\sim
\lambda/t$ by the SO interaction, but the larger Fermi
surface ($|{\bm k}|\sim \lambda/t$) can be
neglected for the zero mode \cite{STF09-2}.
Thus, we concentrate on fermions with ${\bm k}\approx(0,0)$, for which 
${\cal H}^{\rm D}({\bm k})$ 
is decomposed into the following two 2$\times$2 matrices,
\begin{eqnarray}
{\cal H}^{D}_{+}({\bm k})=
\left(
\begin{array}{cc}
\psi_{\rm s}-h &2\lambda (k_y+ik_x) \\
2\lambda (k_y-ik_x) & -\psi_{\rm s}+h
\end{array}
\right), 
\nonumber\\
{\cal H}^{D}_{-}({\bm k})=
\left(
\begin{array}{cc}
\psi_{\rm s}+h &2\lambda (-k_y+ik_x) \\
2\lambda (-k_y-ik_x) & -\psi_{\rm s}-h
\end{array}
\right). 
\label{eq:dual2}
\end{eqnarray}
We consider a single vortex of the ``$p$-wave SF gap'' $2\lambda(\pm k_y+ik_x)$.
The BdG equations for ${\cal H}^{\rm D}_{\pm}$ with the single vortex can be
solved by using the method developed in \cite{read}.
Then, we find a unique zero energy solution with a quasiparticle field
$\gamma^{\dagger}=\int d{\bm r}[u_0\psi^{\dagger}_{+}+v_0\psi_{+}]$,
where $u_0=i(re^{i\theta})^{-1/2}e^{-(h-\psi_{\rm s})r/2\lambda}$, 
$v_0=-i(re^{-i\theta})^{-1/2}e^{-(h-\psi_{\rm s})r/2\lambda}$ \cite{STF09-2}.
The solution is normalizable when (\ref{eq:region}) is satisfied.
This is the Majorana zero energy mode; i.e. $\gamma^{\dagger}=\gamma$.
Using the duality transformation (\ref{eq:duality}),
we found that a vortex in the original Hamiltonian has a single 
Majorana zero mode given by
$D(u_0,0,v_0,0)^{T}$,  
which implies that the vortex is a non-Abelian anyon \cite{read}.

From the construction of the zero mode above, we notice that there is an
important difference between a chiral $p$-wave SF and our system:
While for a spinless chiral $p$-wave SF, a single Majorana zero
mode exists in a vortex of the SF order parameter,
for our non-centrosymmetric $s$-wave SF, a Majorana zero mode exists in a vortex 
twisting a phase of the SO interaction.
This difference can be understood immediately 
from the duality (\ref{eq:dualHamiltonian}) since 
a vortex in a gap function in the dual Hamiltonian is transformed into
a vortex in the SO coupling. 
So far, we have assumed $\epsilon(0,0)=0$, but even when
$\epsilon(0,0)\neq0$, the existence of the non-Abelian topological order
is robust as long as $h$ satisfies (6), because the topological
character is not changed unless the bulk gap closes \cite{SF08}.

For the detection of the non-Abelian anyons, it is desirable
that the zero energy state in a vortex is well separated from
excited states, the interaction with which may cause decoherence.
The excitation energy in the vortex core is due to the kinetic energy
which stems from the derivative term of the BdG equation, 
$2\lambda(\mp i\partial_y+\partial_x)$.
Since the above solution for $(u_0,v_0)$ 
indicates that the size of the vortex core
is $\sim 2\lambda/(h-\psi_s)$,
the excitation energy is of the order
$\sim 2\lambda/(2\lambda/(h-\psi_s))\sim h-\psi_s$.
It can be tuned to be relatively large, and thus
the experimental detection of the non-Abelian anyons is quite feasible.

\begin{figure}
\includegraphics[width=6.5cm]{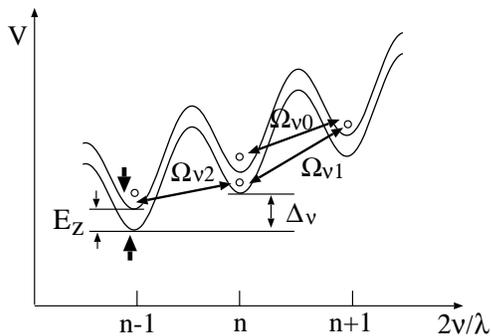}
\caption{\label{fig1}
Setup of the confining optical potential for the ground state. $\nu=x$ or $y$. 
Bold
up and down arrows indicate, respectively, the spin-up and spin-down states.
In this figure, excited levels which mediate the hopping via two Raman lasers
are not shown explicitly.}
\end{figure}

{\it Possible Realization in Cold Fermionic Atoms --}
We now propose an experimental scheme for the realization of
 the topological phase mentioned above in ultracold fermionic atoms. 
It was recently pointed out by several authors that effective
gauge fields interacting with atoms
can be generated by spatially varying laser fields
\cite{ost,ruse,zhu,stan,LCPPPS09}.
These ideas can be utilized for our purpose.
We consider 
fermionic atoms loaded in a 2D periodic optical lattice,
where there is no hopping along the $z$-direction \cite{ost}.
The atoms occupy doubly-degenerate Zeeman levels of the hyperfine 
ground state manifolds, which are, respectively,
the ``spin up'' state 
$|\uparrow\rangle$ and the ``spin down'' state $|\downarrow\rangle$. 
We introduce the Zeeman field to lift the degeneracy.
The Zeeman level split is denoted as $E_{\rm Z}$.
It is assumed that standard tunneling of atoms between sites due to
kinetic energy
is suppressed by the large depth of the optical lattice potential.
Tunneling of atoms between neighboring sites along the $\nu$-direction
($\nu=x,y$)
which conserves spins
is caused by laser beams with the Rabi frequency $\Omega_{\nu 0}$
via optical Raman transitions as proposed in Refs.\cite{ost,jak};
i.e. $\Omega_{\nu 0}=\Omega_{1}'\Omega_{2}'/2\Delta$
with $\Omega_{1,2}'$ the Rabi frequencies for the transition between
the ground state and an excited state, and $\Delta$ detuning from the excited
state.
In addition, tunneling which accompanies spin flip
is also driven by two Raman lasers \cite{ost}.
In Fig.\ref{fig1}, we show the optical lattice setup.
The laser with the Rabi frequency 
$\Omega_{\nu 1}$ ($\Omega_{\nu 2}$) 
is resonant for transition 
$|\uparrow\rangle\rightarrow|\downarrow\rangle $
for the tunneling between neighboring sites in the forward (backward) 
$\nu$-direction. 
As proposed in Ref.\cite{ost}, 
the confining optical potential is tilted along both the $x$-direction 
and the $y$-direction to assure that the forward and backward
tunneling processes are
respectively induced by the lasers with the different Rabi frequencies
$\Omega_{\nu 1}$ and $\Omega_{\nu 2}$,
which are required for the realization of
the Rashba spin-orbit interaction as discussed below.
The energy shift between nearest neighbor sites due to the tilting potential
is $\Delta_x$ for 
the $x$-direction and $\Delta_y$ for the $y$-direction.
We impose the condition $\Delta_x\neq\Delta_y$ 
to prevent tunneling with spin flip along the $y(x)$-direction due to 
the lasers with $\Omega_{x(y)1,2}$.
It is also assumed that the detuning from excited states for optical Raman
transitions is much larger than $\Delta_{x(y)}$, and thus
the spatial variation of the amplitudes of the Rabi frequencies due to
the tilting potential is negligible.
To realize the Rashba spin-orbit interaction for the two Zeeman levels, 
we choose the phases of the lasers as follows.
The lasers are propagating along the $z$-direction with an oscillating factor
$e^{ik_zz}$.
The Rabi frequency $\Omega_{x 2}$ is expressed as 
$\Omega_{x 2}=|\Omega_{x 2}|e^{ik_zz}$.
The phase of the laser $\Omega_{x 1}$ is shifted by $\pi$ 
from that of $\Omega_{x 2}$, and
$\Omega_{x 2}=-\Omega_{x 1}$ holds. 
The phase of $\Omega_{y 1}$ ($\Omega_{y 2}$) is shifted by $-\pi/2$ ($\pi/2$), 
and $\Omega_{y 2}=-i\Omega_{x 1}$, $\Omega_{y 2}=-\Omega_{y 1}$.
Then, the laser-induced tunneling term which accompanies spin flip
is expressed by
$\mathcal{H}_{\rm SO}=\sum_{i}[\lambda_x(c^{\dagger}_{{\bm i}-\hat{x}\downarrow}
c_{{\bm i}\uparrow}-
c^{\dagger}_{{\bm i}+\hat{x}\downarrow}c_{{\bm i}\uparrow})
+i\lambda_y(c^{\dagger}_{{\bm i}-\hat{y}\downarrow}c_{{\bm i}\uparrow}-
c^{\dagger}_{{\bm i}+\hat{y}\downarrow}c_{{\bm i} \uparrow})+{\rm h.c.}]$
with $\lambda_{\nu}=c_{\nu}\int d\mbox{\boldmath $r$}\psi^{*}_{\downarrow}
(\mbox{\boldmath $r$}
-\mbox{\boldmath $r$}_{i-\hat{\mu}})\Omega_{\nu 2}(\mbox{\boldmath $r$})
\psi_{\uparrow}(\mbox{\boldmath $r$}-\mbox{\boldmath $r$}_i)$, 
$\nu=x$, $y$, and
$c_x=1$, $c_y=-i$.
Since we consider the 2D $xy$-plane with $z=0$,
$\lambda_{\nu}$ is real.
For $\lambda_x=\lambda_y$, 
$\mathcal{H}_{\rm SO}$
is the Rashba SO interaction.

To create vortices of the SO interaction, which are key ingredients 
for the realization of non-Abelian anyons, 
we replace the lasers that generate the SO interaction 
with those carrying orbital angular momentum parallel to
the $z$-axis.
Such lasers can be prepared by 
using Laguerre-Gaussian beams \cite{LG}.
Then, a  ``vortex'' of the SO interaction is introduced: 
$\lambda_{\nu}\rightarrow\lambda_{\nu}e^{im\theta}$.
The vorticity $m$ is controlled by changing the configuration of the lasers.
Furthermore, the spatially separated multiple vortices can be generated 
by using the following method.
After introducing a vortex with vorticity $m$ into the system, 
we switch the Laguerre-Gaussian beam to a Gaussian beam, i.e.
a laser without angular momentum \cite{LG}.
The vortex still exists in the system, because of the conservation of
the total angular momentum.
However, for $m>1$, the vortex with higher charges 
become energetically unstable toward 
dissociation into
$m$ vortices with single vorticity, and thus spatially separated 
multiple vortices are created
in the system \cite{SSVPSLPPK04,isoshima}.
This multiple vortex state allows the realization of the non-Abelian
statistics of vortices.

We can use the Feshbach resonance in the $s$-wave channel for the formation of
the $s$-wave Cooper pairs in this system \cite{sbcs,chin}.
Then, 
the topological phase described by the Hamiltonian (\ref{eq:latticehamiltonian}) 
is realized.
In this scheme, the role of the $s$-wave superfluidity is two-fold. One is to
suppress bulk gapless quasiparticles which are harmful for topological stability.
The other one is to generate the superposition of particles and holes, 
which results
in the Majorana quasiparticles in vortices and edge states.
It is noted that the vortices in the SO interaction are stabilized 
by the carefully designed laser setup rather than by macroscopic condensates.
It should be emphasized that 
this experimental scheme is feasible for currently accessible laser techniques.
As mentioned before, 
the non-Abelian anyons are stable for sufficiently low energies
$\ll {\rm min}\{h-\psi_s,\psi_s\}$.
Since $\psi_s$ can be tuned to be large, i.e. $\psi_s\sim E_F$, 
by using the s-wave Feshbach resonance,
the realization of the non-Abelian anyons in this scheme is quite promising.

{\it Summary--} We have proposed a feasible scheme for the realization of
the non-Abelian topological phase in an $s$-wave superfluid of cold atoms
in an optical lattice, in which the non-Abelian anyons exist,
and opens a possible way to realize topological quantum computation.

The authors thank the organizers of the symposium, ``Topological Aspects
of Solid State Physics'', at YITP, Kyoto, where this work has been
 started.
This work was partly supported by the Grant-in-Aids for
the Global COE Program
``The Next Generation of Physics, Spun from Universality and Emergence''
and Scientific Research 
(Grant No.18540347, Grant No.19014009, Grant No.19052003) from MEXT of Japan.

\bibliography{top-atom}

\begin{thebibliography}{39}
\expandafter\ifx\csname natexlab\endcsname\relax\def\natexlab#1{#1}\fi
\expandafter\ifx\csname bibnamefont\endcsname\relax
  \def\bibnamefont#1{#1}\fi
\expandafter\ifx\csname bibfnamefont\endcsname\relax
  \def\bibfnamefont#1{#1}\fi
\expandafter\ifx\csname citenamefont\endcsname\relax
  \def\citenamefont#1{#1}\fi
\expandafter\ifx\csname url\endcsname\relax
  \def\url#1{\texttt{#1}}\fi
\expandafter\ifx\csname urlprefix\endcsname\relax\def\urlprefix{URL }\fi
\providecommand{\bibinfo}[2]{#2}
\providecommand{\eprint}[2][]{\url{#2}}

\bibitem[{\citenamefont{Wen and Niu}(1990)}]{wen}
\bibinfo{author}{\bibfnamefont{X.~G.} \bibnamefont{Wen}} \bibnamefont{and}
  \bibinfo{author}{\bibfnamefont{Q.}~\bibnamefont{Niu}},
  \bibinfo{journal}{Phys.\ Rev.\ B} \textbf{\bibinfo{volume}{41}},
  \bibinfo{pages}{9377} (\bibinfo{year}{1990}).

\bibitem[{\citenamefont{Moore and Read}(1991)}]{moore}
\bibinfo{author}{\bibfnamefont{G.}~\bibnamefont{Moore}} \bibnamefont{and}
  \bibinfo{author}{\bibfnamefont{N.}~\bibnamefont{Read}},
  \bibinfo{journal}{Nucl.\ Phys.} \textbf{\bibinfo{volume}{B360}},
  \bibinfo{pages}{362} (\bibinfo{year}{1991}).

\bibitem[{\citenamefont{Nayak and Wilczek}(1996)}]{nayak}
\bibinfo{author}{\bibfnamefont{C.}~\bibnamefont{Nayak}} \bibnamefont{and}
  \bibinfo{author}{\bibfnamefont{F.}~\bibnamefont{Wilczek}},
  \bibinfo{journal}{Nucl.\ Phys.} \textbf{\bibinfo{volume}{B479}},
  \bibinfo{pages}{529} (\bibinfo{year}{1996}).

\bibitem[{\citenamefont{Murakami et~al.}(2004)\citenamefont{Murakami, Nagaosa,
  and Zhang}}]{zhang2}
\bibinfo{author}{\bibfnamefont{S.}~\bibnamefont{Murakami}},
  \bibinfo{author}{\bibfnamefont{N.}~\bibnamefont{Nagaosa}}, \bibnamefont{and}
  \bibinfo{author}{\bibfnamefont{S.~C.} \bibnamefont{Zhang}},
  \bibinfo{journal}{Phys.\ Rev.\ Lett.} \textbf{\bibinfo{volume}{93}},
  \bibinfo{pages}{156804} (\bibinfo{year}{2004}).

\bibitem[{\citenamefont{Kane and Mele}(2005)}]{kane}
\bibinfo{author}{\bibfnamefont{C.~L.} \bibnamefont{Kane}} \bibnamefont{and}
  \bibinfo{author}{\bibfnamefont{E.~J.} \bibnamefont{Mele}},
  \bibinfo{journal}{Phys.\ Rev.\ Lett.} \textbf{\bibinfo{volume}{95}},
  \bibinfo{pages}{146802} (\bibinfo{year}{2005}).

\bibitem[{\citenamefont{Bernevig and Zhang}(2006)}]{zhang}
\bibinfo{author}{\bibfnamefont{B.~A.} \bibnamefont{Bernevig}} \bibnamefont{and}
  \bibinfo{author}{\bibfnamefont{S.~C.} \bibnamefont{Zhang}},
  \bibinfo{journal}{Phys.\ Rev.\ Lett.} \textbf{\bibinfo{volume}{96}},
  \bibinfo{pages}{106802} (\bibinfo{year}{2006}).

\bibitem[{\citenamefont{Kitaev}(2006)}]{kitaev}
\bibinfo{author}{\bibfnamefont{A.}~\bibnamefont{Kitaev}},
  \bibinfo{journal}{Ann.\ Phys.} \textbf{\bibinfo{volume}{321}},
  \bibinfo{pages}{2} (\bibinfo{year}{2006}).

\bibitem[{\citenamefont{Lee et~al.}(2007)\citenamefont{Lee, Zhang, and
  Xiang}}]{lee}
\bibinfo{author}{\bibfnamefont{D.~H.} \bibnamefont{Lee}},
  \bibinfo{author}{\bibfnamefont{G.~M.} \bibnamefont{Zhang}}, \bibnamefont{and}
  \bibinfo{author}{\bibfnamefont{T.}~\bibnamefont{Xiang}},
  \bibinfo{journal}{Phys.\ Rev.\ Lett.} \textbf{\bibinfo{volume}{99}},
  \bibinfo{pages}{196805} (\bibinfo{year}{2007}).

\bibitem[{\citenamefont{Read and Green}(2000)}]{read}
\bibinfo{author}{\bibfnamefont{N.}~\bibnamefont{Read}} \bibnamefont{and}
  \bibinfo{author}{\bibfnamefont{D.}~\bibnamefont{Green}},
  \bibinfo{journal}{Phys.\ Rev.\ B} \textbf{\bibinfo{volume}{61}},
  \bibinfo{pages}{10267} (\bibinfo{year}{2000}).

\bibitem[{\citenamefont{Ivanov}(2001)}]{ivanov}
\bibinfo{author}{\bibfnamefont{D.~A.} \bibnamefont{Ivanov}},
  \bibinfo{journal}{Phys.\ Rev.\ Lett.} \textbf{\bibinfo{volume}{86}},
  \bibinfo{pages}{268} (\bibinfo{year}{2001}).

\bibitem[{\citenamefont{Stern et~al.}(2004)\citenamefont{Stern, von Oppen, and
  Mariani}}]{stern2}
\bibinfo{author}{\bibfnamefont{A.}~\bibnamefont{Stern}},
  \bibinfo{author}{\bibfnamefont{F.}~\bibnamefont{von Oppen}},
  \bibnamefont{and} \bibinfo{author}{\bibfnamefont{E.}~\bibnamefont{Mariani}},
  \bibinfo{journal}{Phys.\ Rev.\ B} \textbf{\bibinfo{volume}{70}},
  \bibinfo{pages}{205338} (\bibinfo{year}{2004}).

\bibitem[{\citenamefont{Tsutsumi et~al.}(2008)\citenamefont{Tsutsumi, Kawakami,
  Mizushima, Ichioka, and Machida}}]{machida}
\bibinfo{author}{\bibfnamefont{Y.}~\bibnamefont{Tsutsumi}},
  \bibinfo{author}{\bibfnamefont{T.}~\bibnamefont{Kawakami}},
  \bibinfo{author}{\bibfnamefont{T.}~\bibnamefont{Mizushima}},
  \bibinfo{author}{\bibfnamefont{M.}~\bibnamefont{Ichioka}}, \bibnamefont{and}
  \bibinfo{author}{\bibfnamefont{K.}~\bibnamefont{Machida}},
  \bibinfo{journal}{Phys.\ Rev.\ Lett.} \textbf{\bibinfo{volume}{101}},
  \bibinfo{pages}{135302} (\bibinfo{year}{2008}).

\bibitem[{\citenamefont{Micheli et~al.}(2006)\citenamefont{Micheli, Brennen,
  and Zoller}}]{zol}
\bibinfo{author}{\bibfnamefont{A.}~\bibnamefont{Micheli}},
  \bibinfo{author}{\bibfnamefont{G.~K.} \bibnamefont{Brennen}},
  \bibnamefont{and} \bibinfo{author}{\bibfnamefont{P.}~\bibnamefont{Zoller}},
  \bibinfo{journal}{Nature\ Physics} \textbf{\bibinfo{volume}{2}},
  \bibinfo{pages}{341} (\bibinfo{year}{2006}).

\bibitem[{\citenamefont{Sato}(2003)}]{Sato03}
\bibinfo{author}{\bibfnamefont{M.}~\bibnamefont{Sato}},
  \bibinfo{journal}{Phys.\ Lett.} \textbf{\bibinfo{volume}{B575}},
  \bibinfo{pages}{126} (\bibinfo{year}{2003}).

\bibitem[{\citenamefont{Kitaev}(2003)}]{kitaev2}
\bibinfo{author}{\bibfnamefont{A.}~\bibnamefont{Kitaev}},
  \bibinfo{journal}{Ann. Phys.} \textbf{\bibinfo{volume}{303}},
  \bibinfo{pages}{2} (\bibinfo{year}{2003}).

\bibitem[{\citenamefont{Stone and Chung}(2006)}]{stone1}
\bibinfo{author}{\bibfnamefont{M.}~\bibnamefont{Stone}} \bibnamefont{and}
  \bibinfo{author}{\bibfnamefont{S.-B.} \bibnamefont{Chung}},
  \bibinfo{journal}{Phys.\ Rev.\ B} \textbf{\bibinfo{volume}{73}},
  \bibinfo{pages}{014505} (\bibinfo{year}{2006}).

\bibitem[{\citenamefont{Freedman et~al.}(2003)\citenamefont{Freedman, Kitaev,
  Larsen, and Wang}}]{freedman}
\bibinfo{author}{\bibfnamefont{M.}~\bibnamefont{Freedman}},
  \bibinfo{author}{\bibfnamefont{A.}~\bibnamefont{Kitaev}},
  \bibinfo{author}{\bibfnamefont{M.}~\bibnamefont{Larsen}}, \bibnamefont{and}
  \bibinfo{author}{\bibfnamefont{Z.}~\bibnamefont{Wang}},
  \bibinfo{journal}{Bull.\ Amer.\ Math.\ Soc.} \textbf{\bibinfo{volume}{40}},
  \bibinfo{pages}{31} (\bibinfo{year}{2003}).

\bibitem[{\citenamefont{Tewari et~al.}(2007)\citenamefont{Tewari, Sarma, Nayak,
  Zhang, and Zoller}}]{tewari2}
\bibinfo{author}{\bibfnamefont{S.}~\bibnamefont{Tewari}},
  \bibinfo{author}{\bibfnamefont{S.~D.} \bibnamefont{Sarma}},
  \bibinfo{author}{\bibfnamefont{C.}~\bibnamefont{Nayak}},
  \bibinfo{author}{\bibfnamefont{C.}~\bibnamefont{Zhang}}, \bibnamefont{and}
  \bibinfo{author}{\bibfnamefont{P.}~\bibnamefont{Zoller}},
  \bibinfo{journal}{Phys.\ Rev.\ Lett.} \textbf{\bibinfo{volume}{98}},
  \bibinfo{pages}{010506} (\bibinfo{year}{2007}).

\bibitem[{\citenamefont{Osterloh et~al.}(2005)\citenamefont{Osterloh, Baig,
  Santos, Zoller, and Lewenstein}}]{ost}
\bibinfo{author}{\bibfnamefont{K.}~\bibnamefont{Osterloh}},
  \bibinfo{author}{\bibfnamefont{M.}~\bibnamefont{Baig}},
  \bibinfo{author}{\bibfnamefont{L.}~\bibnamefont{Santos}},
  \bibinfo{author}{\bibfnamefont{P.}~\bibnamefont{Zoller}}, \bibnamefont{and}
  \bibinfo{author}{\bibfnamefont{M.}~\bibnamefont{Lewenstein}},
  \bibinfo{journal}{Phys.\ Rev.\ Lett.} \textbf{\bibinfo{volume}{95}},
  \bibinfo{pages}{010403} (\bibinfo{year}{2005}).

\bibitem[{\citenamefont{Ruseckas et~al.}(2005)\citenamefont{Ruseckas,
  Juzeliunas, Ohberg, and Fleischhauer}}]{ruse}
\bibinfo{author}{\bibfnamefont{J.}~\bibnamefont{Ruseckas}},
  \bibinfo{author}{\bibfnamefont{G.}~\bibnamefont{Juzeliunas}},
  \bibinfo{author}{\bibfnamefont{P.}~\bibnamefont{Ohberg}}, \bibnamefont{and}
  \bibinfo{author}{\bibfnamefont{M.}~\bibnamefont{Fleischhauer}},
  \bibinfo{journal}{Phys.\ Rev.\ Lett.} \textbf{\bibinfo{volume}{95}},
  \bibinfo{pages}{010404} (\bibinfo{year}{2005}).

\bibitem[{\citenamefont{Zhu et~al.}(2006)\citenamefont{Zhu, Fu, Wu, Zhang, and
  Duan}}]{zhu}
\bibinfo{author}{\bibfnamefont{S.~L.} \bibnamefont{Zhu}},
  \bibinfo{author}{\bibfnamefont{H.}~\bibnamefont{Fu}},
  \bibinfo{author}{\bibfnamefont{C.~J.} \bibnamefont{Wu}},
  \bibinfo{author}{\bibfnamefont{S.~C.} \bibnamefont{Zhang}}, \bibnamefont{and}
  \bibinfo{author}{\bibfnamefont{L.~M.} \bibnamefont{Duan}},
  \bibinfo{journal}{Phys.\ Rev.\ Lett.} \textbf{\bibinfo{volume}{97}},
  \bibinfo{pages}{240401} (\bibinfo{year}{2006}).

\bibitem[{\citenamefont{Stanescu et~al.}(2007)\citenamefont{Stanescu, Zhang,
  and Galitski}}]{stan}
\bibinfo{author}{\bibfnamefont{T.~D.} \bibnamefont{Stanescu}},
  \bibinfo{author}{\bibfnamefont{C.}~\bibnamefont{Zhang}}, \bibnamefont{and}
  \bibinfo{author}{\bibfnamefont{V.}~\bibnamefont{Galitski}},
  \bibinfo{journal}{Phys.\ Rev.\ Lett.} \textbf{\bibinfo{volume}{99}},
  \bibinfo{pages}{110403} (\bibinfo{year}{2007}).

\bibitem[{\citenamefont{Lin et~al.}(2009)\citenamefont{Lin, Compton, Perry,
  Phillips, Porto, and Spielman}}]{LCPPPS09}
\bibinfo{author}{\bibfnamefont{Y.~J.} \bibnamefont{Lin}},
  \bibinfo{author}{\bibfnamefont{R.~L.} \bibnamefont{Compton}},
  \bibinfo{author}{\bibfnamefont{A.~R.} \bibnamefont{Perry}},
  \bibinfo{author}{\bibfnamefont{W.~D.} \bibnamefont{Phillips}},
  \bibinfo{author}{\bibfnamefont{J.~V.} \bibnamefont{Porto}}, \bibnamefont{and}
  \bibinfo{author}{\bibfnamefont{I.~B.} \bibnamefont{Spielman}},
  \bibinfo{journal}{Phys.\ Rev.\ Lett.} \textbf{\bibinfo{volume}{102}},
  \bibinfo{pages}{130401} (\bibinfo{year}{2009}).

\bibitem[{\citenamefont{Sato}(2006)}]{Sato06}
\bibinfo{author}{\bibfnamefont{M.}~\bibnamefont{Sato}},
  \bibinfo{journal}{Phys.\ Rev.\ B} \textbf{\bibinfo{volume}{73}},
  \bibinfo{pages}{214502} (\bibinfo{year}{2006}).

\bibitem[{\citenamefont{Sato and Fujimoto}(2009)}]{SF08}
\bibinfo{author}{\bibfnamefont{M.}~\bibnamefont{Sato}} \bibnamefont{and}
  \bibinfo{author}{\bibfnamefont{S.}~\bibnamefont{Fujimoto}},
  \bibinfo{journal}{Phys.\ Rev.\ B} \textbf{\bibinfo{volume}{79}},
  \bibinfo{pages}{094504} (\bibinfo{year}{2009}).

\bibitem[{\citenamefont{Tanaka et~al.}(2009)\citenamefont{Tanaka, Yokoyama,
  Balatsky, and Nagaosa}}]{TYBN08}
\bibinfo{author}{\bibfnamefont{Y.}~\bibnamefont{Tanaka}},
  \bibinfo{author}{\bibfnamefont{T.}~\bibnamefont{Yokoyama}},
  \bibinfo{author}{\bibfnamefont{A.~V.} \bibnamefont{Balatsky}},
  \bibnamefont{and} \bibinfo{author}{\bibfnamefont{N.}~\bibnamefont{Nagaosa}},
  \bibinfo{journal}{Phys.\ Rev.\ B} \textbf{\bibinfo{volume}{79}},
  \bibinfo{pages}{060505} (\bibinfo{year}{2009}).

\bibitem[{\citenamefont{Zhang et~al.}(2008)\citenamefont{Zhang, Tewari,
  Lutchyn, and Sarma}}]{ZTLDS}
\bibinfo{author}{\bibfnamefont{C.}~\bibnamefont{Zhang}},
  \bibinfo{author}{\bibfnamefont{S.}~\bibnamefont{Tewari}},
  \bibinfo{author}{\bibfnamefont{R.~M.} \bibnamefont{Lutchyn}},
  \bibnamefont{and} \bibinfo{author}{\bibfnamefont{S.~D.} \bibnamefont{Sarma}},
  \bibinfo{journal}{Phys.\ Rev.\ Lett.} \textbf{\bibinfo{volume}{101}},
  \bibinfo{pages}{160401} (\bibinfo{year}{2008}).

\bibitem[{\citenamefont{Inada et~al.}(2008)\citenamefont{Inada, Horikoshi,
  Nakajima, Kuwata-Gonokami, Ueda, and Mukaiyama}}]{inada}
\bibinfo{author}{\bibfnamefont{M.}~\bibnamefont{Inada}},
  \bibinfo{author}{\bibfnamefont{M.}~\bibnamefont{Horikoshi}},
  \bibinfo{author}{\bibfnamefont{S.}~\bibnamefont{Nakajima}},
  \bibinfo{author}{\bibfnamefont{M.}~\bibnamefont{Kuwata-Gonokami}},
  \bibinfo{author}{\bibfnamefont{M.}~\bibnamefont{Ueda}}, \bibnamefont{and}
  \bibinfo{author}{\bibfnamefont{T.}~\bibnamefont{Mukaiyama}},
  \bibinfo{journal}{Phys.\ Rev.\ Lett.} \textbf{\bibinfo{volume}{101}},
  \bibinfo{pages}{100401} (\bibinfo{year}{2008}).

\bibitem[{\citenamefont{Bourdel et~al.}(2004)\citenamefont{Bourdel, Khaykovich,
  Cubizolles, Zhang, Chevy, Teichmann, Tarruell, Kokkelmans, and
  Salomon}}]{sbcs}
\bibinfo{author}{\bibfnamefont{T.}~\bibnamefont{Bourdel}},
  \bibinfo{author}{\bibfnamefont{L.}~\bibnamefont{Khaykovich}},
  \bibinfo{author}{\bibfnamefont{J.}~\bibnamefont{Cubizolles}},
  \bibinfo{author}{\bibfnamefont{J.}~\bibnamefont{Zhang}},
  \bibinfo{author}{\bibfnamefont{F.}~\bibnamefont{Chevy}},
  \bibinfo{author}{\bibfnamefont{M.}~\bibnamefont{Teichmann}},
  \bibinfo{author}{\bibfnamefont{L.}~\bibnamefont{Tarruell}},
  \bibinfo{author}{\bibfnamefont{S.~J. J. M.~F.} \bibnamefont{Kokkelmans}},
  \bibnamefont{and} \bibinfo{author}{\bibfnamefont{C.}~\bibnamefont{Salomon}},
  \bibinfo{journal}{Phys.\ Rev.\ Lett.} \textbf{\bibinfo{volume}{93}},
  \bibinfo{pages}{050401} (\bibinfo{year}{2004}).

\bibitem[{\citenamefont{Chin et~al.}(2006)\citenamefont{Chin, Miller, Liu,
  Stan, Setiawan, Sanner, Xu, and Ketterle}}]{chin}
\bibinfo{author}{\bibfnamefont{J.~K.} \bibnamefont{Chin}},
  \bibinfo{author}{\bibfnamefont{D.~E.} \bibnamefont{Miller}},
  \bibinfo{author}{\bibfnamefont{Y.}~\bibnamefont{Liu}},
  \bibinfo{author}{\bibfnamefont{C.}~\bibnamefont{Stan}},
  \bibinfo{author}{\bibfnamefont{W.}~\bibnamefont{Setiawan}},
  \bibinfo{author}{\bibfnamefont{C.}~\bibnamefont{Sanner}},
  \bibinfo{author}{\bibfnamefont{K.}~\bibnamefont{Xu}}, \bibnamefont{and}
  \bibinfo{author}{\bibfnamefont{W.}~\bibnamefont{Ketterle}},
  \bibinfo{journal}{Nature} \textbf{\bibinfo{volume}{443}},
  \bibinfo{pages}{961} (\bibinfo{year}{2006}).

\bibitem[{\citenamefont{Rashba}(1960)}]{rash}
\bibinfo{author}{\bibfnamefont{E.~I.} \bibnamefont{Rashba}},
  \bibinfo{journal}{Sov.\ Phys.\ Solid State} \textbf{\bibinfo{volume}{2}},
  \bibinfo{pages}{1109} (\bibinfo{year}{1960}).

\bibitem[{\citenamefont{Thouless et~al.}(1982)\citenamefont{Thouless, Kohmoto,
  Nightingale, and den Nijs}}]{TKNN}
\bibinfo{author}{\bibfnamefont{D.~J.} \bibnamefont{Thouless}},
  \bibinfo{author}{\bibfnamefont{M.}~\bibnamefont{Kohmoto}},
  \bibinfo{author}{\bibfnamefont{M.~P.} \bibnamefont{Nightingale}},
  \bibnamefont{and} \bibinfo{author}{\bibfnamefont{M.}~\bibnamefont{den Nijs}},
  \bibinfo{journal}{Phys.\ Rev.\ Lett.} \textbf{\bibinfo{volume}{49}},
  \bibinfo{pages}{405} (\bibinfo{year}{1982}).

\bibitem[{\citenamefont{Fujimoto}(2007)}]{fuji3}
\bibinfo{author}{\bibfnamefont{S.}~\bibnamefont{Fujimoto}},
  \bibinfo{journal}{J.\ Phys.\ Soc.\ Jpn.} \textbf{\bibinfo{volume}{76}},
  \bibinfo{pages}{051008} (\bibinfo{year}{2007}).

\bibitem[{\citenamefont{Frigeri et~al.}(2004)\citenamefont{Frigeri, Agterberg,
  Koga, and Sigrist}}]{Frigeri}
\bibinfo{author}{\bibfnamefont{P.~A.} \bibnamefont{Frigeri}},
  \bibinfo{author}{\bibfnamefont{D.~F.} \bibnamefont{Agterberg}},
  \bibinfo{author}{\bibfnamefont{A.}~\bibnamefont{Koga}}, \bibnamefont{and}
  \bibinfo{author}{\bibfnamefont{M.}~\bibnamefont{Sigrist}},
  \bibinfo{journal}{Phys.\ Rev.\ Lett.} \textbf{\bibinfo{volume}{92}},
  \bibinfo{pages}{097001} (\bibinfo{year}{2004}).

\bibitem[{\citenamefont{Sato et~al.}()\citenamefont{Sato, Takahashi, and
  Fujimoto}}]{STF09-2}
\bibinfo{author}{\bibfnamefont{M.}~\bibnamefont{Sato}},
  \bibinfo{author}{\bibfnamefont{Y.}~\bibnamefont{Takahashi}},
  \bibnamefont{and} \bibinfo{author}{\bibfnamefont{S.}~\bibnamefont{Fujimoto}},
  \eprint{in preparation}.

\bibitem[{\citenamefont{Jaksch and Zoller}(2003)}]{jak}
\bibinfo{author}{\bibfnamefont{D.}~\bibnamefont{Jaksch}} \bibnamefont{and}
  \bibinfo{author}{\bibfnamefont{P.}~\bibnamefont{Zoller}},
  \bibinfo{journal}{New\ J.\ Phys.} \textbf{\bibinfo{volume}{5}},
  \bibinfo{pages}{56.1} (\bibinfo{year}{2003}).

\bibitem[{\citenamefont{Anderson et~al.}(2006)\citenamefont{Anderson, Ryu,
  Clad\'e, Natarajan, Vaziri, Helmerson, and Phillips}}]{LG}
\bibinfo{author}{\bibfnamefont{M.~F.} \bibnamefont{Anderson}},
  \bibinfo{author}{\bibfnamefont{C.}~\bibnamefont{Ryu}},
  \bibinfo{author}{\bibfnamefont{P.}~\bibnamefont{Clad\'e}},
  \bibinfo{author}{\bibfnamefont{V.}~\bibnamefont{Natarajan}},
  \bibinfo{author}{\bibfnamefont{A.}~\bibnamefont{Vaziri}},
  \bibinfo{author}{\bibfnamefont{K.}~\bibnamefont{Helmerson}},
  \bibnamefont{and} \bibinfo{author}{\bibfnamefont{W.~D.}
  \bibnamefont{Phillips}}, \bibinfo{journal}{Phys.\ Rev.\ Lett.}
  \textbf{\bibinfo{volume}{97}}, \bibinfo{pages}{170406}
  (\bibinfo{year}{2006}).

\bibitem[{\citenamefont{Isoshima et~al.}(2007)\citenamefont{Isoshima, Okano,
  Yasuda, Kasa, Huhtamaki, Kumakura, and Takahashi}}]{isoshima}
\bibinfo{author}{\bibfnamefont{T.}~\bibnamefont{Isoshima}},
  \bibinfo{author}{\bibfnamefont{M.}~\bibnamefont{Okano}},
  \bibinfo{author}{\bibfnamefont{H.}~\bibnamefont{Yasuda}},
  \bibinfo{author}{\bibfnamefont{K.}~\bibnamefont{Kasa}},
  \bibinfo{author}{\bibfnamefont{J.~A.~M.} \bibnamefont{Huhtamaki}},
  \bibinfo{author}{\bibfnamefont{M.}~\bibnamefont{Kumakura}}, \bibnamefont{and}
  \bibinfo{author}{\bibfnamefont{Y.}~\bibnamefont{Takahashi}},
  \bibinfo{journal}{Phys.\ Rev.\ Lett.} \textbf{\bibinfo{volume}{99}},
  \bibinfo{pages}{200403} (\bibinfo{year}{2007}).

\bibitem[{\citenamefont{Shin et~al.}(2004)\citenamefont{Shin, Saba,
  Vengalattore, Pasquini, Sanner, Leanhardt, Prentiss, Pritchard, and
  Ketterle}}]{SSVPSLPPK04}
\bibinfo{author}{\bibfnamefont{Y.}~\bibnamefont{Shin}},
  \bibinfo{author}{\bibfnamefont{M.}~\bibnamefont{Saba}},
  \bibinfo{author}{\bibfnamefont{M.}~\bibnamefont{Vengalattore}},
  \bibinfo{author}{\bibfnamefont{T.~A.} \bibnamefont{Pasquini}},
  \bibinfo{author}{\bibfnamefont{C.}~\bibnamefont{Sanner}},
  \bibinfo{author}{\bibfnamefont{A.~E.} \bibnamefont{Leanhardt}},
  \bibinfo{author}{\bibfnamefont{M.}~\bibnamefont{Prentiss}},
  \bibinfo{author}{\bibfnamefont{D.~E.} \bibnamefont{Pritchard}},
  \bibnamefont{and} \bibinfo{author}{\bibfnamefont{W.}~\bibnamefont{Ketterle}},
  \bibinfo{journal}{Phys.\ Rev.\ Lett.} \textbf{\bibinfo{volume}{93}},
  \bibinfo{pages}{160406} (\bibinfo{year}{2004}).

\end{thebibliography}

\end{document}